\documentclass[onecolumn,showpacs,pre,amsmath,amssymb,nofootnoteinbib]{revtex4-1}

\usepackage{graphicx}
\usepackage{dcolumn}
\usepackage{bm}
\usepackage{color}

\def\b{\mathbf}

\usepackage{graphicx}%
\usepackage{dcolumn}%
\usepackage{bm}%
\usepackage{amsbsy}
\usepackage{amssymb}
\usepackage{amsmath}
\usepackage{graphics, color}
\usepackage[normalem]{ulem}
\usepackage[usenames,dvipsnames,svgnames,table]{xcolor}

\usepackage{amssymb,amsfonts,amsmath}

\usepackage{amssymb,amsfonts,amsmath}
\def\b{\mathbf}

\DeclareMathOperator{\sech}{sech}

\begin{document}


\title{Geometrically controlled snapping transitions in shells with curved creases}


\author{
Nakul P. Bende$^{1}$}
\thanks{these two authors contributed equally}
\author{
Arthur A. Evans$^{2}$}
\thanks{these two authors contributed equally}
\author{
Sarah Innes-Gold$^1$}
\author{
Luis A. Marin$^3$}
\author{
Itai Cohen$^4$}
\author{Ryan C. Hayward$^1$}
\email{hayward@umass.edu}
\author{Christian D. Santangelo$^2$}
\email{csantang@physics.umass.edu}

\affiliation{$^1$ Department of Polymer Science and Engineering, University of Massachusetts, Amherst}
\affiliation{$^2$ Department of Physics, University of Massachusetts, Amherst}
\affiliation{$^3$ Department of Mechanical and Industrial Engineering, University of Massachusetts, Amherst}
\affiliation{$^4$ Department of Physics, Cornell University}

\begin{abstract}
Curvature and mechanics are intimately connected for thin materials, and this coupling between geometry and physical properties is readily seen in folded structures from intestinal villi and pollen grains, to wrinkled membranes and programmable metamaterials. While the well-known rules and mechanisms behind folding a flat surface have been used to create deployable structures and shape transformable materials, folding of curved shells is still not fundamentally understood. Curved shells naturally deform by simultaneously bending and stretching, and while this coupling gives them great stability for engineering applications, it makes folding a surface of arbitrary curvature a non-trivial task. Here we discuss the geometry of folding a creased shell, and demonstrate theoretically the conditions under which it may fold continuously. When these conditions are violated we show, using experiments and simulations, that shells undergo rapid snapping motion to fold from one stable configuration to another. Although material asymmetry is a proven mechanism for creating this bifurcation of stability, for the case of a creased shell, the inherent geometry itself serves as a barrier to folding. We discuss here how two fundamental geometric concepts, creases and curvature, combine to allow rapid transitions from one stable state to another. Independent of material system and length scale, the design rule that we introduce here explains how to generate snapping transitions in arbitrary surfaces, thus facilitating the creation of programmable multi-stable materials with fast actuation capabilities.
\end{abstract}
%
%
\maketitle

Curved shells are generally used to enhance structural stability \cite{vaziri2008localized,vella2012indentation,lazarus2012geometry}, since the coupling between bending and stretching makes them energetically costly to deform. The consequences of this coupling are seen in both naturally occurring scenarios, such as intestinal villi and pollen grains \cite{shyer2013villification,katifori2010foldable}, and finds use in man-made structures such as programmable meta-materials \cite{schenk2013geometry,wei2013geometric,jesse2014muira}. When these shells have  multiple stable configurations, geometrically enhanced rigidity produces an energetic barrier dominated by stretching energy, often over a relatively small range of deformation, which leads to the high forces and rapid acceleration usually associated with a ``snap-through" transition\cite{forterre2005venus,crosby2007lens,hayashi2009mechanics,smith2011elastic,shankar2013contactless,lu2014charge,pandey2014}. For example, Venus flytraps (\textit{Dionaea muscipula}) use this mechanism to generate their leaf snapping motion\cite{forterre2005venus}, hummingbirds (\textit{Aves: Trochilidae}) twist and rotate their curved beaks to catch insect prey\cite{smith2011elastic} , and engineered micro lenses use a combination of bending and stretching energy to rapidly switch from convex to concave shapes in order to tune their optical properties \cite{crosby2007lens}. Despite the ability to engineer these snapping transitions in a variety of material systems, there does not exist a general geometric design rule for generating snapping transitions between stable states of arbitrary surfaces. This stands in stark contrast to folding of a flat sheet, which has well known rules and consequences \cite{guest1994deployable,tachi2009generalization,demaine2011reconstructing}. In these cases, weakening the material locally by introducing a crease allows the sheet to deform without stretching, and thus access globally isometric states without requiring nonlinear deformations. 





\section{Folding a shell}

Inspired by ideas from origami, we discuss here the folding of curved surfaces with creases. We show, both theoretically and experimentally, that the curvature of this crease can be used to control the continuity of the deformation. The proposed design principle arises purely from geometry, and does not rely on special material systems or anisotropy to generate rapid snap-through transitions.

\begin{figure*}[h]
\includegraphics[width=\textwidth]{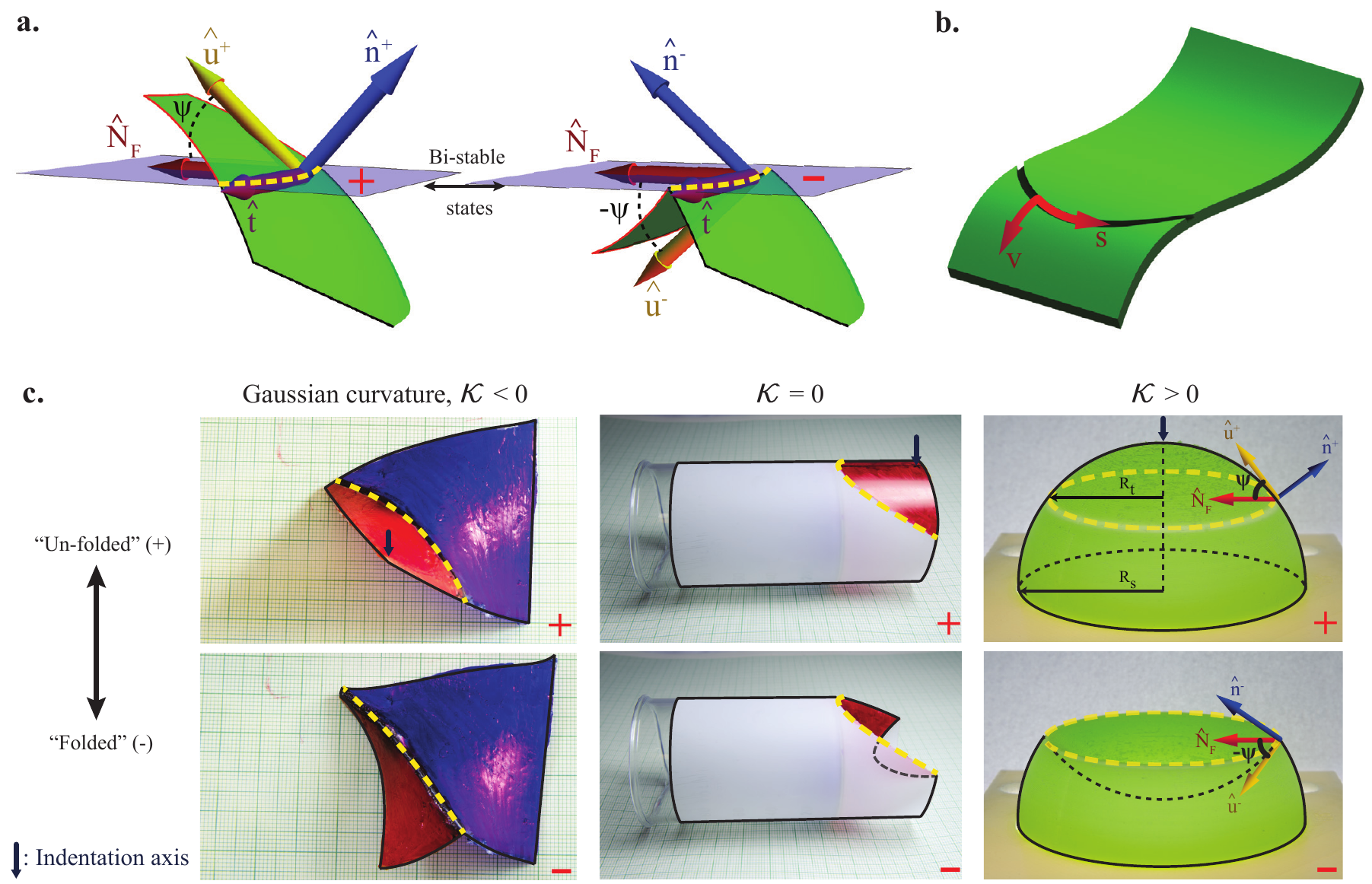}
\centering
\caption{
\textbf{Discontinuities in folding curved shells} \textbf{a.} Natural and folded states of a creased shell, denoted as the \textbf{+} (unfolded) and \textbf{-} (folded) conformations.  Tangent and normal vectors to each surface $\pm$ are given by $\b{\hat{u}^\pm}$ and ($\b{\hat{n}^\pm}$), respectively. $\b{\hat{N}_F}$ is the normal to the curve, while $\b{\hat{t}}$ indicates tangent to both the crease and the surface. \textbf{b.} Creasing shells involves thinning the shell locally along a curve that lies on the surface. Definition of a local coordinate system $\{s,v\}$ on crease facilitates theoretical analysis.\textbf{c.} Experimental sample with bi-stable states of a creased (i) helicoid ($K<0$) made from poly(caprolactone), (ii) cylinder ($K=0$) with rigid boundary conditions made from a poly(ethylene terephthalate) sheet and (iii) spherical shell ($K>0$) of thickness $t$, radius $R_s$ with a crease radius $R_t$ made from poly(vinyl siloxane). Load-displacement assays are done by indenting these shells along the axis pictured (black arrow). Creases and sample boundaries are marked with a dotted yellow and black curves respectively. 
}
\label{fig:geometry}
\end{figure*}

We first consider introducing a crease onto a non-Euclidean shell with finite Gaussian curvature $\mathcal{K}$ (Fig. \ref{fig:geometry}a). Though this concept has been realized on rare occasions in art, the continuum mechanics of a creased shell is far from fully understood. Origami, and deployable structures in general, utilize isometric deformations of thin sheets by introducing local weakening (i.e. creasing) of the material. Likewise for shells, including a region of locally thin material (Fig. \ref{fig:geometry}b) allows deformations that are locally isometric; whether or not this deformation is continuous is governed by the geometric relations we derive here.

Using the geometric quantities defined in Fig. \ref{fig:geometry}a (see Methods), we derive a relationship between the crease and the surface parameterized by the angle $\psi$:

\begin{gather}
\psi=\pm \cos^{-1}\left(\sqrt{1-(\kappa_N/\kappa)^2}\right),
\end{gather}
where $\kappa$ is the curvature of the crease and $\kappa_N\equiv-\kappa\b{\hat{N}}_F\cdot\b{\hat{n}}=\kappa\sin\psi$ is the normal curvature. This relation states that, in addition to its unbent configuration, any surface can be folded along this crease into an un-stretched state provided that the angle $\psi\rightarrow -\psi$ (Fig. \ref{fig:geometry}a, right). This ``mirror reflection" folding does not stretch the surface, but under what conditions can the surface be folded continuously?


\subsection{Normal curvature controls continuity of folding}

Considering isometric deformations, the only contribution to the elastic energy density of folding comes from bending energy $\mathcal{E}_B\sim B (H^\pm)^2$, where the mean curvature of the surface is $H^\pm=\frac{1}{2}Tr\{\mathcal{II^\pm}/\mathcal{I}\}$ and $B$ is the bending rigidity of the shell (see Methods). For surfaces of normal curvature zero, these shells may be folded continuously. However, for $\kappa_N\neq0$, as the surface is folded the normal curvature on one side of the crease must pass through zero, and we find that the mean curvature $H^\pm\sim1/\tan\psi$. In order to fold the surface through $\psi=0$ the surface must develop corners or kinks along the crease so that the mean curvature diverges, similar to stress focusing phenomena seen in other curved surfaces \cite{arroyo2003nonlinear,witten2007stress,pauchard1998contact,vaziri2008localized,vaziri2009mechanics,nasto2013localization,datta2012delayed}. Since bending energy scales as the mean curvature squared, the barrier between the folded and unfolded state is infinite for isometric deformations. In any real material, as the shell bends the energy will reach a scale where stretching becomes favorable. Thus, geometry alone gives a prediction for a design rule: by introducing a crease with finite $\kappa_N$ an energetic barrier is created between a pair of locally isometric, bi-stable states. For any finite thickness shell, such transitions will require stretching of the surface and often lead to violent snaps.
\section{Experimental results}

\begin{figure}[h]
\includegraphics[width=0.4\textwidth]{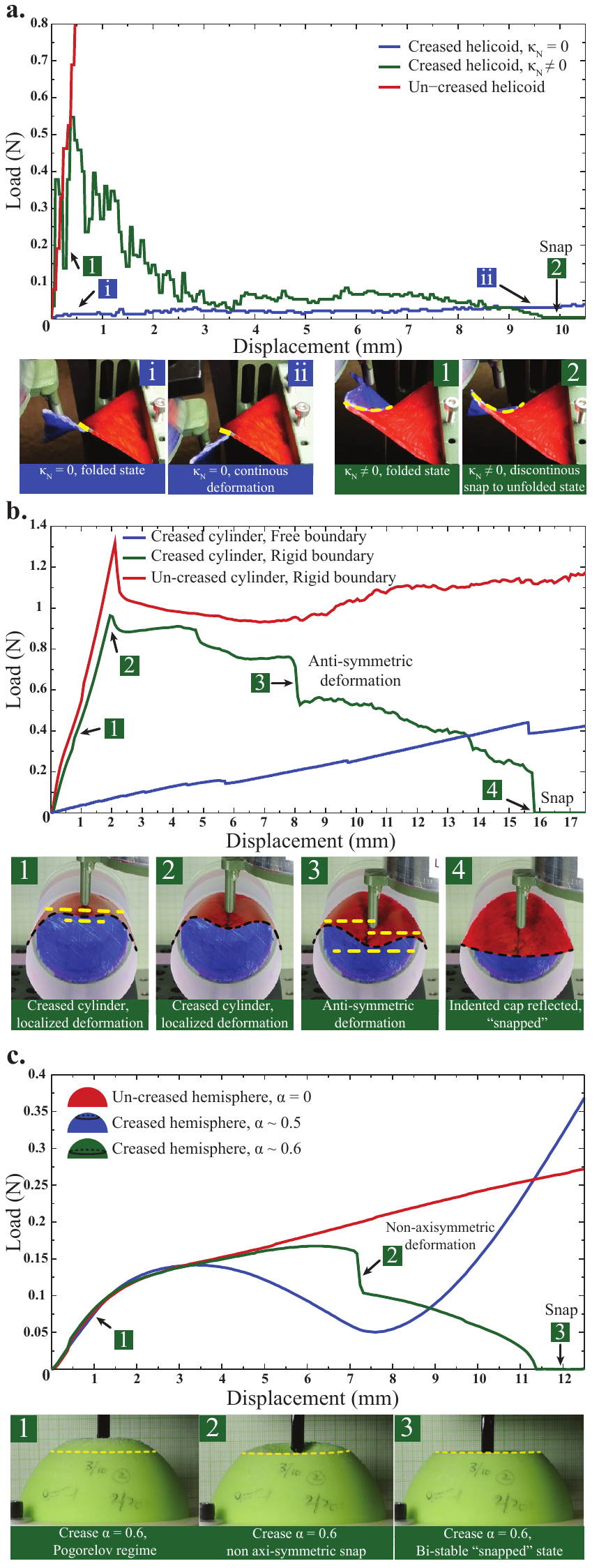}
\caption{
\textbf{Force-displacement characteristics of creased shells} \textbf{a)} Force-displacement for a helicoid: in the absence of a crease, the response is relatively rigid (red). For creases with zero $\kappa_N$ there is no snap (blue), while for creases with $\kappa_N\neq 0$, a snap is clearly visible as the force curve drops to zero (green) (see movie M1). \textbf{b)} Force-displacement for an uncreased cylinder (red), a creased cylinder with free boundary (blue), and a creased cylinder with rigid boundary (green). Since the cylinder can bend without stretching, boundary conditions  heavily influence the snapping behavior (see movie M2). \textbf{c)} Spherical shells with crease radius $\alpha = R_t/R_s = 0$ (red), $0.5$ (blue), and $0.6$ (green). For the smaller crease, no stable snap is observed, while for the larger crease a non-axisymmetric deformation, followed by a stable snap, occurs under indentation (see movie M3).
}
\label{fig:force}
\end{figure}

To test the applicability of this design rule we crease elastomeric and plastic helicoids, cylinders and spheres, whose  Gaussian curvatures $\mathcal{K}$ are respectively negative, zero and positive (Fig. \ref{fig:geometry}c, Methods). The measurements consist of deforming the surface with a point indenter and measuring force-displacement response during folding of the shell. For most creases on these prototypical surfaces, we observe snapping behavior where the force drops to zero as the surface snaps and loses contact with the indenter. However, for creases with $\kappa_N=0$, the force-displacement curve rises monotonically and no snap occurs, as predicted by our design rule.

\subsection{Negative Gaussian curvature, helicoid}

To place a crease of zero $\kappa_N$ on a surface, the surface must have non-positive Gaussian curvature. On a helicoid, examples of such creases are the straight ``construction lines" that are used to generate the surface. For these cases the fold angle can be changed continuously and no snapping transition is observed (\ref{fig:force}a, insets i and ii). For deformation along any other arbitrary crease with finite $\kappa_N$, we observe a snap (Fig. \ref{fig:geometry}c, \ref{fig:force}a insets 1 and 2, Movie M1). 

\subsection{Zero Gaussian curvature, cylinder}

Creases with finite $\kappa_N$ on a cylinder can be created by intersecting the surface with a plane at an oblique angle, as shown in Fig. \ref{fig:geometry}c. According to our hypothesis, we expect the cylinder to undergo a snapping transition when deformed along this crease. Remarkably, despite the introduction of a crease, the cylinder displays global bending deformation instead of snapping, in line with results from other studies on un-creased cylinders \cite{boudaoud2000dynamics,vaziri2008localized,vaziri2009mechanics} (Fig. \ref{fig:force}b). Such global deformations arise since cylinders have $\mathcal{K}=0$, and thus can easily bend without stretching, so that there is a pathway accessible to the shell that costs less energy than snapping but still satisfies the constraint imposed by the indenter. These pathways can be eliminated by imposing rigid boundary conditions on the free ends of the cylinder by inserting a rigid cylindrical plug. Without access to these bending modes, the cylinder undergoes a snapping transition that first involves an antisymmetric mode, followed by a full snap (Fig. \ref{fig:force}b, insets, Movie M2). 

\subsection{Positive Gaussian curvature, sphere}

Mechanisms involving pure bending are avoided in spherical shells, since surfaces with $\mathcal{K}>0$ naturally require stretching for almost any deformation of the surface. Thus we expect that intersecting a spherical surface with a plane to create a crease (Fig. \ref{fig:geometry}c) with finite $\kappa_N$ will result in a snap. We create hemispheres with different crease radius $R_t$ and sphere radius of curvature $R_s$, and define the normalized crease radius as $\alpha = R_t/R_s$. The force displacement curve for these surfaces are shown in (Fig. \ref{fig:force}c). For an un-creased shell ($\alpha = 0$), we observe a monotonically increasing behavior, similar to previous studies\cite{gupta1999axial, vaziri2008localized, nasto2013localization}. Surprisingly, we find that for small $\alpha$ the load displacement curve develops a local minimum, but the ``snapped'' state remains unstable. For higher values of $\alpha$ the folding pathway leads first to an unstable, non-axisymmetric snap, soon followed by a well-defined stable snap (Fig. \ref{fig:force}c, insets, Movie M3).

\begin{figure*}[h]
\includegraphics[width=\textwidth]{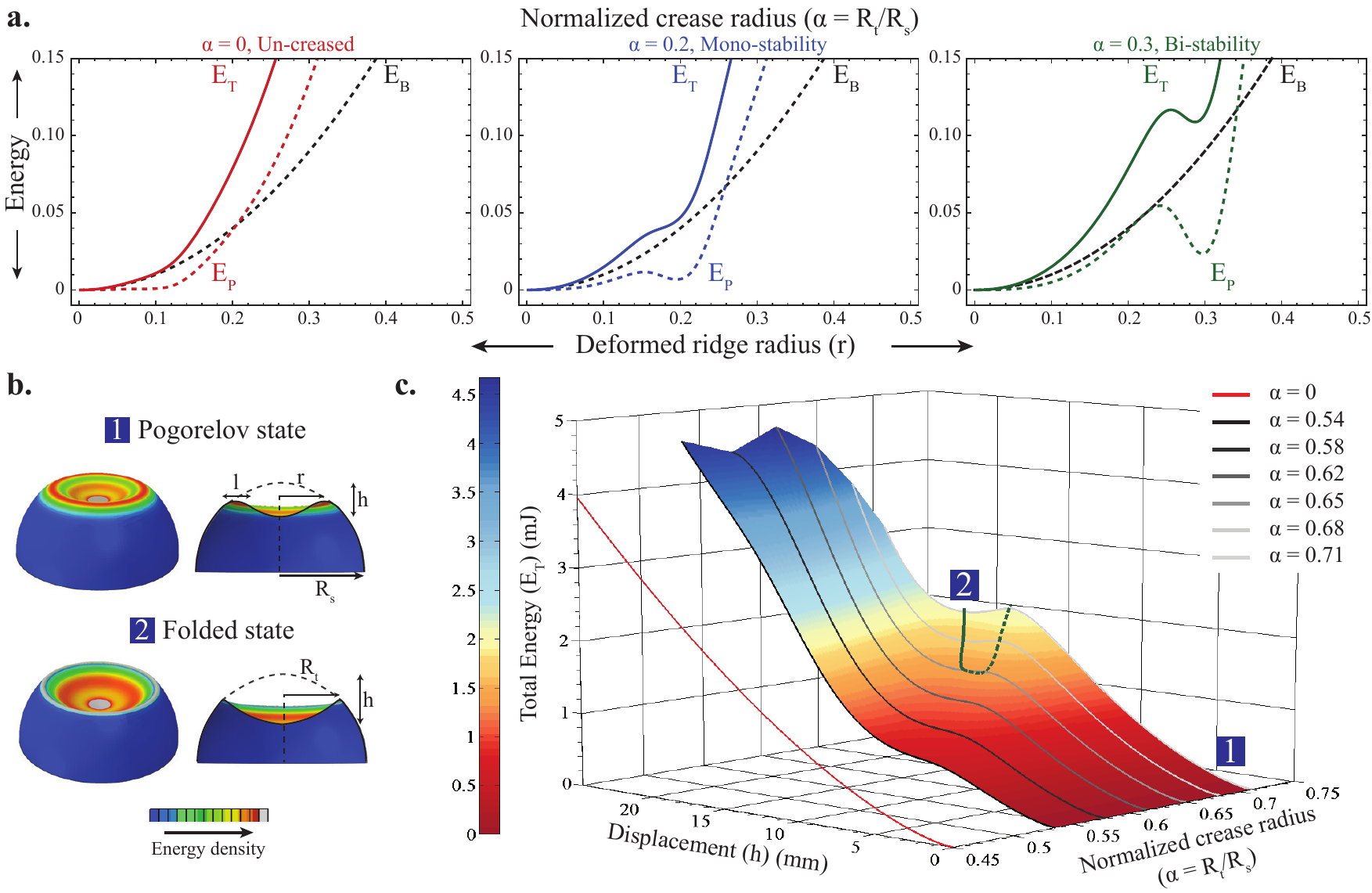}
\centering
\caption{
\textbf{Energy landscape of creased spheres} \textbf{a.} For an un-creased shell, the energy of an indented shell is composed of the bending energy $E_B$ (black dotted) and the Pogorelov ridge $E_P$ (colored dotted). For a creased shell $E_P$ takes a substantial dip at $r\sim R_t$, but the total energy $E_T$ (colored solid) only has a local minimum if the crease is large enough. \textbf{b.} Schematics for (1) Pogorelov state of a deformed spherical shell, with representative ridge (of size $\sim l$) at a radius $r$ and (2) folded state of a creased spherical shell, with a crease radius $R_t$ \textbf{c.} Numerically calculated energy landscape for a creased shell with $\gamma\approx 10^4$ for a variety of $\alpha$. The Pogorelov solution is recovered for $\alpha=0$ (red plot), while for small values of $\alpha$ the energy gain from crease is insufficient to create a local minimum. However, above a critical $\alpha$, local minima (solid green) and maxima (dashed green) bifurcate to generate a region of stability.
}
\label{fig:energy}
\end{figure*}

To understand how stability depends on $\alpha$, we consider the classical problem of indenting a spherical shell (Fig. \ref{fig:energy}a,b). There is a well-known nearly isometric deformation of a sphere, the Pogorelov ridge, which is seen for displacements larger than the thickness but smaller than the crease size \cite{landau1959course}. This deformation regime is characterized by an inverted bulge of radius $r$ and bounded by a ridge of size $\ell\sim\sqrt{t R_s}$ (Fig. \ref{fig:energy}b). The energy for this state has a bending energy contribution from the inverted bulge that scales as $E_B\sim \mathcal{B} (r/R_s)^2$, while the Pogorelov ridge, which contains all the stretching energy, scales as $E_P\sim Y (t/R_s)^{5/2} r^3$, with $Y$ the Young's modulus of the material. Hence the total energy for deformation scaled by $\mathcal{B}$ can be expressed as

\begin{gather}
\frac{E_P+E_B}{\mathcal{B}}\approx \left(\frac{r}{R_s}\right)^2+15\left((1-\nu^2)\gamma\right)^{1/4}\left(\frac{r}{R_s}\right)^3,
\end{gather}
where $\gamma$ is the F\"{o}ppl-von K\'{a}rm\'{a}n number $\gamma\equiv \hat{Y}R_s^2/\mathcal{B}$, with $\hat{Y}$ as the stretching modulus of the material. For a thin shell $\hat{Y}=Yt$, and $\mathcal{B}=Yt^3/12(1-\nu^2)$, where $\nu$ is Poisson's ratio, such that $\gamma\sim(R_s\sqrt{12(1-\nu^2)}/t)^2$. The F\"{o}ppl-von K\'{a}rm\'{a}n number characterizes the balance between bending and stretching energies, and can be defined even for structures that are not technically thin shells, such as viruses and polymerized membranes \cite{lidmar2003virus}.

In the case of a creased sphere, we assume that the deformation of the shell retains the same structure as this classical solution, but now the thickness of the sphere in the Pogorelov ridge is a function of the bulge radius $r$. Since the crease is localized to a small region on the surface (Fig. \ref{fig:force}b), we define $t\equiv t_0 g(r-R_t)$, where $t_0$ is the regular thickness of the shell and the function $g(x)$ is of the generic localized form such that $g(0)=\epsilon\ll 1$ and $g(x\rightarrow\infty)=1$.  With this assumption, the energy for the modified Pogorelov solution becomes

\begin{gather}
\frac{E_P+E_B}{\mathcal{B}_0}\approx\left(\frac{r}{R_s}\right)^2+15(1-\nu^2)^{1/4}\gamma^{1/4}g^{5/2}\left(\frac{r}{R_s}\right)^3,
\end{gather}
where $\mathcal{B}_0=Yt_0^3/12(1-\nu^2)$. For schematic purposes, we choose $g(r)=(1-\epsilon)\sech^2{(\frac{r-R_t}{t_0})}$, where $\epsilon<1$ is the fractional change in the thickness, and plot these results with $\epsilon=1/2$ for various different crease radii in Fig. \ref{fig:energy}a. There is now a local minimum in the Pogorelov energy centered at $r=R_t$ that generates an energy barrier that competes with bending energy. We estimate the regime of stability by setting the size of this energy barrier equal to the bending energy contained in the folded state:

\begin{gather}
\label{scaling}
\gamma^{1/4}\left(\frac{r^*}{R_s}\right)^3\sim\left(\frac{R_t}{R_s}\right)^2
\end{gather} 

The balance of these two energies yields a critical Pogorelov radius $r^*$ such that a local minimum will appear at a finite value of $r$. If $r^*<R_t$, this bi-stability is accessible to the shell. Solving Eq. (\ref{scaling}) for $r^*$, and setting $r^*\equiv R_t$ yields the scaling law 

\begin{gather}
\frac{1}{\sqrt{\gamma}}\sim \alpha^2.
\end{gather}

%


To confirm this hypothesis, we use finite element simulations (ABAQUS, \textit{Dassault Systemes}) to determine the conditions under which there is a stable snap. For linearly elastic materials this system is fully characterized by two dimensionless numbers, the reduced crease radius $\alpha$ and the F\"{o}ppl-von K\'{a}rm\'{a}n number $\gamma$. We report the total energy for axisymmetric solutions with $\gamma=10^4$ (corresponding to the elastomeric hemispheres discussed here), as a function of the indenter displacement ($h$) and the normalized crease radius ($\alpha$)  in Fig. \ref{fig:energy}c. We find that, beyond a critical crease radius, there is a bifurcation of stability and the energy curves develop a well-defined local minimum (solid) and maximum (dashed), with the region between these curves denoting a basin of attraction for the folded state. 

\begin{figure}[h]
\includegraphics[width=0.45\textwidth]{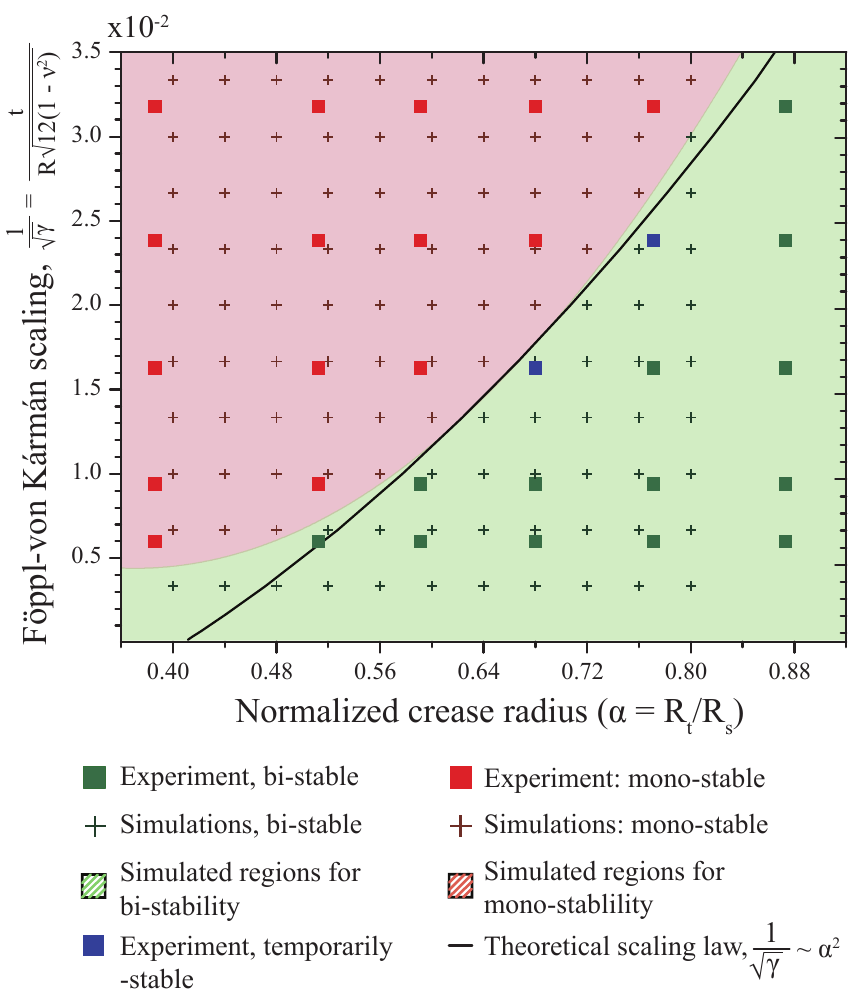}
\centering
\caption{
\textbf{Phase diagram for spherical shells} Phase diagram for snapping behavior of spherical shells over a wide range of geometrical parameters $\alpha$ and $1/\sqrt{\gamma}$. Stability behavior in experiments is characterized as bi-stable (\textcolor{OliveGreen}{$\blacksquare$}, switches to folded state through a snapping mechanism), mono-stable (\textcolor{Crimson}{$\blacksquare$}, prefers unfolded state) or temporarily-stable (\textcolor{Navy}{$\blacksquare$}, closer to phase boundary: snap back on a time-scale of seconds without external perturbations). Finite element simulations (points solved denoted with \textbf{+}) provide regions of mono-stability (red shading) and bi-stability (green shading) and the theoretical scaling law agrees well with both experiments and simulations. Each experimental data point was analyzed for at least 3 shells of appropriate parameters. 
}
\label{phase}
\end{figure}

By examining creased hemispherical shells over a wide range of $\gamma$, we identify how the critical crease radius necessary for generating a stable snap depends on shell thickness (Fig. \ref{phase}).  Through numerical simulations, we find that for increasing thickness larger values of the crease radius are required to create a stable snap. Moreover, the boundary between mono- and bi-stable regimes is well-captured by the scaling argument suggested by our modified Pogorelov ridge calculation. As a final test of our theory and simulations, we conduct a series of experiments on spherical shells with a range of $\gamma$ and $\alpha$, and identify the stability of the folded state. These too reveal a boundary between bi-stability and mono-stability, that is in excellent agreement with our numerical calculations and theoretical law.  Moreover, for some samples we observe the presence of folded states that are temporarily stable (for times on the order of seconds); the proximity of these samples to the predicted phase boundaries further demonstrates the agreement between experiment, simulation, and theory.

\section{Conclusion}
The ability to introduce tunable bi-stability into a curved shell via structural inhomogeneity represents a major step in generating programmable materials with rapid actuation capabilities. While inhomogeneous shells have already been predicted to serve as a template for constructing tunable shapes \cite{paulose2013buckling}, and used to design next generation substances such as lock-and-key colloids \cite{sacanna2010lock} or controllably collapsible capsules \cite{datta2012delayed}, our geometric design principle adds further insight into controlling the mechanics of thin shells. Since the speed of the snap arises from stretching in the shell, inertia mediates the transition at the speed of sound in the material (see movies M1-M3), and crucially, the snap is unimpeded by poro-elasticity or hydraulic damping as displayed in many natural snapping systems \cite{maha2005limits}. Our work lays the foundation for developing non-Euclidean origami, in which multiple folds and vertices combine to create new structures. Finally, since the principles we describe in this paper are purely geometric, they open the door for developing design paradigms independent of length-scale and material system.


\section{Materials and methods}

\subsection{Shell fabrication}

3-dimensional models of different geometries were designed in a CAD software. The non-Euclidean geometries (helicoid and hemisphere) were fabricated using a commercial 3d printer (\textit{uDimensions, Stratys Inc.}) to obtain two-part molds with embossed features to generate creases. The hemispherical shells were fabricated using poly(vinyl siloxane) by curing a commercially available two part base-catalyst mixture  (\textit{Zhermack SpA Elite Double 32}, Elastic modulus $(Y) \sim 100$ kPa). Prior to filling the mold, the 1.2:1 base:catalyst mixture was degassed for removing bubbles which may defect the sample. Helicoid samples were fabricated using poly(caprolactone) (\textit{Monomer-Polymer \& Dajac Labs, 1258}, $Y = 353$ MPa), by first melting the polymer, filling and cooling down inside the molds. The hemispherical and helicoid shells studied were 1 mm thick, and the crease had a rectangular cross-section 0.75 mm deep ($\epsilon = 0.75$) and 1mm wide along the appropriate curve. Only the samples without any bubbles/ other structural defects were included for testing. Owing to their Euclidean nature, cylinders could be fabricated using a conventional two dimensional technique. Here, we used a commercial laser cutter (\textit{Zing Epilog 16}) to score a poly(ethylene terephthalate) sheet (\textit{Grafix Dura-Lar®}, 120 $\mu$m thick, $Y\sim 5$ GPa) with a curve. The shape of this plane curve is set to be sinusoidal such that when the sheet is wrapped to form a cylinder, the resulting space curve is the intersection between a plane and a cylinder at an oblique angle.

\subsection{Load displacement characterization}

A custom-built force displacement device, combining a linear translation stage (\textit{Zaber Technologies Inc., T-LSM 100}) and a load cell (\textit{Loadstar Sensors Inc., RPG-10}), was used to perform strain controlled force measurements. Point indenters (radius ratio of indenter with respect to shell $\sim 0.05$) were 3D printed, and samples were indented at a strain rate of 5 mm/min. Data collection and analysis was performed using an in-house algorithm in MATLAB (\textit{The Mathworks}), without any signal processing/ filtering. 

\subsection{Geometry of folding an arbitrary surface}

The crease is characterized by a space curve that lies entirely on the surface of the shell (Fig. \ref{fig:geometry}a), parametrized by a tangent vector $\b{\hat{t}}$. The derivative of the tangent along the curve defines a vector whose magnitude is the curvature \textit{$\kappa$}, and whose direction is given by the Frenet unit normal $\b{\hat{N}}_F$. At any point on the surface, the vector $\b{\hat{u}}$ that is simultaneously tangent to the surface and orthogonal to $\b{\hat{t}}$ makes an angle \textit{$\psi$} with the Frenet normal (Fig. \ref{fig:geometry}a, left), while the normal to the surface is defined by the vector $\b{\hat{n}}$. The surface itself is composed of two regions that are divided by the crease, each parametrized by a local orthonormal frame. In a frame of reference where one surface is fixed in space a local orthonormal frame $\{\b{\hat{t}},\b{\hat{u}}^+,\b{\hat{n}}^+\}$ defines the two surfaces in the unfolded state. When the surface is folded, another frame $\{\b{\hat{t}},\b{\hat{u}}^-,\b{\hat{n}}^-\}$ is used to signify the change from the undeformed state. 

To determine the allowed isometric shapes of a doubly-curved shell, we define a metric tensor on the surface, using the crease to define a local coordinate system $\{s,v\}$ (Fig. \ref{fig:geometry}b). $s$ is an arc length coordinate that runs tangent to the crease, $v$ a measure that runs perpendicular to the crease. From this, we find the first fundamental form of the surface (i.e. the metric):
\begin{gather}
\mathcal{I}=dv^2+\rho(v,s)^2ds^2,
\end{gather}
where $\rho(v,s)$ is related to the curvature of the surface. Close to the crease $v\approx0$, and it is straightforward to find the components of the second fundamental form $\mathcal{II}^\pm=N^\pm ds^2+2M^\pm dsdv+L^\pm dv^2$, where $\pm$ in this case denotes the two sides of the fold. Obtaining these components requires the use of Gauss's {\it Theorema Egregium} (that is, $\det\mathcal{II}/\det\mathcal{I}=\mathcal{K}$) to impose the condition of isometry on the surfaces, and we find that, when $\kappa_N\neq0$,
\begin{gather}
N^\pm=\pm\kappa_N\\
M^\pm=\pm\partial_s\psi+\tau\\
L^\pm=\frac{1}{\kappa_N}\left(\mathcal{K}+(\pm\partial_s\psi+\tau)^2\right),
\end{gather}
where $\tau$ is the torsion of the crease. These equations hold for $\kappa_N\neq 0$, but for $\kappa_N=0$ we find that the bending energy does not diverge and thus the shell may be folded continuously. Conversely, for $\kappa_N\neq0$, the bending energy diverges, thus indicating an energy barrier between folded states. 

\begin{acknowledgments}
\subsection{Acknowledgements}
The authors thank Jesse L. Silverberg, Thomas C. Hull, Douglas P. Holmes, and Dominic Vella for illuminating discussions. We are grateful to Michael J. Imburgia, Alfred J. Crosby, Mindy Dai, Sam R. Nugen for assistance with both the 3D printer and laser cutter, and to Pedro Reis for discussions regarding the fundamentals of shell mechanics and insight on elastomeric design. This work was funded by the National Science Foundation through EFRI ODISSEI-1240441 with additional support to S.I.-G. through the UMass MRSEC DMR-0820506 REU program.
\end{acknowledgments}

\subsection{Author Contributions} N.P.B. and A.A.E. contributed equally to this work. N.P.B., A.A.E., R.C.H. and C.D.S. designed the research. R.C.H., C.D.S. and I.C. supervised this work. N.P.B., A.A.E., S.I.G., L.A.M., R.C.H. and C.D.S. contributed towards fabrication of experimental samples. N.P.B. and R.C.H. designed the custom load displacement setup, and carried out characterization. A.A.E. carried out numerical simulations. All authors contributed in writing of the manuscript.

\end{document}